\documentclass [aps,prx,twocolumn,superscriptaddress]{revtex4-1}

\usepackage{graphicx}
\usepackage{amsmath}
\usepackage{amsmath, amsfonts, bm, color, ulem}
\newcommand{\uvec} [1]{\boldsymbol{\widehat{\textbf{#1}}}}
\begin{document}


\title{Experimental evidence of spin-orbit torque from metallic interfaces}


\author{A. Anad\'on}
\email[]{alberto.anadon@imdea.org, paolo.perna@imdea.org}
\affiliation{IMDEA Nanociencia, c/ Faraday 9, 28049 Madrid, Spain.}
\author{R. Guerrero}
\affiliation{IMDEA Nanociencia, c/ Faraday 9, 28049 Madrid, Spain.}

\author{J. A. Jover-Galtier}
\affiliation{Centro Universitario de la Defensa de Zaragoza, Academia General Militar, ctra. de Huesca s/n. 50090 Zaragoza, Spain}
\affiliation{Instituto de Biocomputación y Física de Sistemas Complejos \& Departamento de Física Teórica, Universidad de Zaragoza, 50009 Zaragoza, Spain}
\author{A. Gud\'in}
\affiliation{IMDEA Nanociencia, c/ Faraday 9, 28049 Madrid, Spain.}

\author{J. M. D\'iez}
\affiliation{IMDEA Nanociencia, c/ Faraday 9, 28049 Madrid, Spain.}
\affiliation{Departamento de Física de la Materia Condensada \& Departamento de Física Aplicada \& Instituto Nicolás Cabrera, Universidad Autónoma de Madrid, 28049 Madrid, Spain.}

\author{P. Olleros-Rodr\'iguez}
\affiliation{IMDEA Nanociencia, c/ Faraday 9, 28049 Madrid, Spain.}

\author{R. Miranda}
\affiliation{IMDEA Nanociencia, c/ Faraday 9, 28049 Madrid, Spain.}
\affiliation{Departamento de Física de la Materia Condensada \& Departamento de Física Aplicada \& Instituto Nicolás Cabrera, Universidad Autónoma de Madrid, 28049 Madrid, Spain.}
\affiliation{IFIMAC, Universidad Autónoma de Madrid, 28049 Madrid, Spain.}

\author{J. Camarero}
\affiliation{IMDEA Nanociencia, c/ Faraday 9, 28049 Madrid, Spain.}
\affiliation{Departamento de Física de la Materia Condensada \& Departamento de Física Aplicada \& Instituto Nicolás Cabrera, Universidad Autónoma de Madrid, 28049 Madrid, Spain.}
\affiliation{IFIMAC, Universidad Autónoma de Madrid, 28049 Madrid, Spain.}
\author{P. Perna$^*$}
\affiliation{IMDEA Nanociencia, c/ Faraday 9, 28049 Madrid, Spain.}
\email[]{paolo.perna@imdea.org}

\date{\today}
\begin{abstract}
Spin currents can modify the magnetic state of ferromagnetic ultrathin films through spin-orbit torque. They may be generated by means of  spin-orbit interaction  by either bulk or interfacial phenomena. Electrical transport measurements reveal a six-fold increase of the spin-orbit torque accompanied by a drastic reduction of the spin Hall magnetoresistance upon the introduction of a Cu interlayer in a Pt/Cu/Co/Pt structure with perpendicular magnetic anisotropy. We analyze the dependence of the spin Hall magnetoresistance with the thickness of the interlayer in the frame of a drift diffusion model that provides information on the expected spin currents and spin accumulations in the system. The results demonstrate that the major responsible of both effects is spin memory loss at the interface. The enhancement of the spin-orbit  torque when introducing an interlayer opens the possibility to design more efficient spintronic devices based on materials that are cheap and abundant such as copper.
\end{abstract}

\maketitle

\section{introduction}
The latest advances in spintronics are based on structures with perpendicular magnetic anisotropy \cite{Mangin2006,Ikeda2010,Baek2018} in which the spin currents are used to produce changes in the magnetization of a magnetic layer \cite{MihaiMiron2010,Avci2015,MihaiMiron2014,Garello2013}. This effect is known as spin-orbit torque (SOT) and can be enhanced by suitably engineering multilayer stacks composed by alternated magnetic/non-magnetic metals \cite{Garello2013,Zhu2019a,Gambardella2011,Avci2017,Saitoh2006,avci2019effects}. The typical structures employed to manipulate the magnetization via SOT are multilayers whose basic constituent is a ferromagnetic (FM) layer adjacent to heavy metal(s) (HM), which confer large spin-orbit coupling and promote the perpendicular magnetic anisotropy.

The spin-orbit coupling is responsible for bulk phenomena such as the spin Hall effect, the conversion of an electric current into a transverse pure spin current, and its counterpart, the inverse spin Hall effect. 
At interfaces, where the structural inversion symmetry is broken, interfacial spin orbit interactions are expected to play a critical role in controlling the electronic states as well as the magnetization configurations \cite{Baek2018}.
The lack of inversion symmetry, proximity effects, orbital hybridization, charge/spin dependent transport, etc; make interfaces between different materials a unique playground environment for the observation of novel physical phenomena.

\begin{figure*}
    \includegraphics [trim=0 1cm 0 0,width=\textwidth]{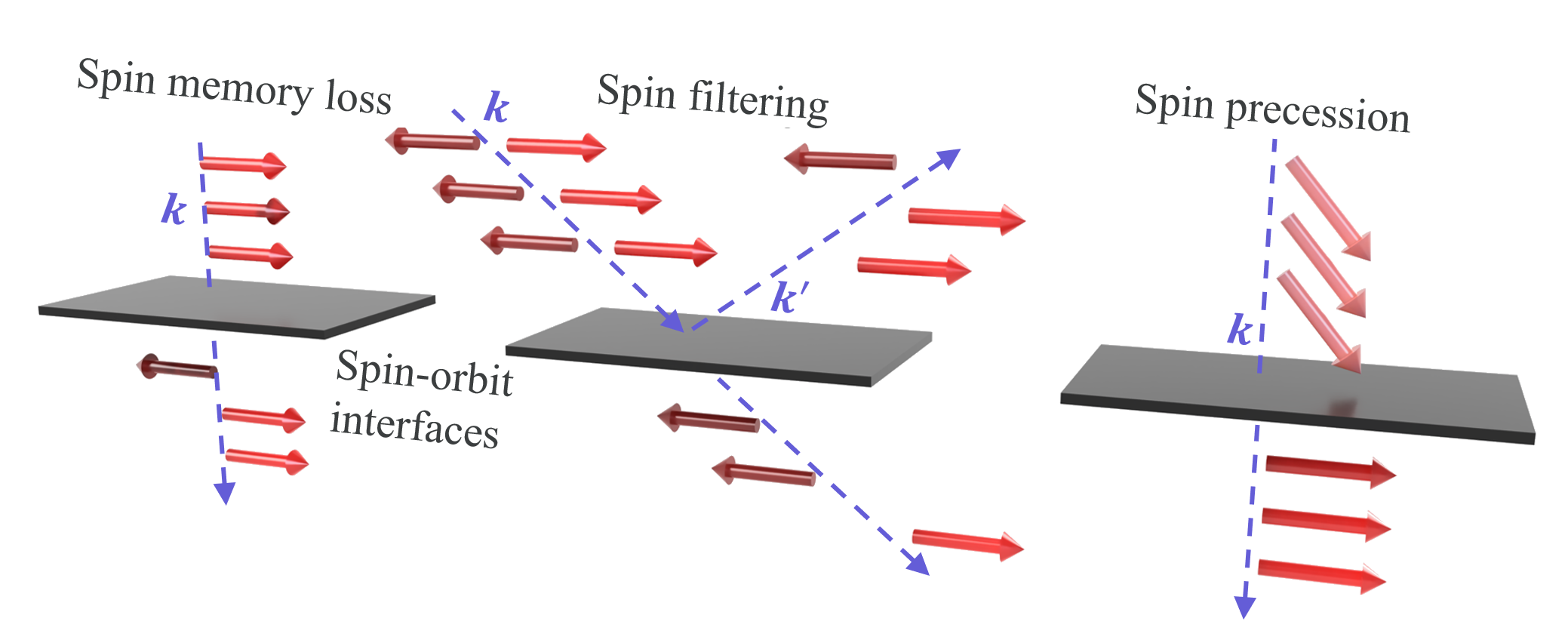}
    \caption{(Color online) {\bf Schematics of spin currents associated phenomena at interfaces with spin-orbit interaction}. \textit{Spin memory loss} is a decoherence process that leads to a reduction of the spin polarization when a spin current passes through an interface. \textit{Spin-orbit filtering} happens due to carriers with spins parallel and antiparallel to the interfacial spin-orbit field experiencing different scattering amplitudes. When this occurs, reflected and transmitted carriers are spin polarized even if incoming carriers are unpolarized. \textit{Spin precession} originates from the interaction of electrons with the interfacial spin-orbit field while they traverse the interface and then scatter off \cite{Amin2018}. These phenomena can generate spin-orbit torques.}
    \label{fig1}
\end{figure*}

In presence of a spin current travelling through an interface \cite{Amin2016,Amin2016a,Zhu2019}, both bulk and interfacial spin-orbit interactions contribute to the SOT. For instance, sizeable spin-Hall effect has been observed in materials with negligible intrinsic spin-orbit coupling (like Cu or Al, or even insulating oxides) \cite{fujiwara20135}, revealing the relevance of the interfacial effects.


At the interfaces, SOT may be in fact produced by spin precession, spin filtering and spin memory loss \cite{Rojas-Sanchez2013,Amin2018,Dolui2017} (schematically illustrated in figure \ref{fig1}), which have all the effect to unbalance the spin accumulation and to produce a torque on the magnetization of an adjacent FM layer. However, the microscopic mechanisms that drive the current-induced magnetization dynamics are still controversial. In fact, the first in-plane current induced magnetization switching in a FM was attributed initially to an interfacial effect (Rashba-induced spin-orbit interaction), \cite{Miron2011} while later was argued to be induced by a bulk spin-Hall effect \cite{Liu2012a}.

To separate bulk and interfacial contributions, asymmetric interfaces, in which the FM layer is embedded between two different non-magnetic (NM) or HM materials, can be used to unbalance the spin-orbit interactions originated at the two interfaces.  Recently, an enhancement of the spin-orbit torque has been observed in interfacial spin scattering experiments by introducing sub-monolayer thick metal interlayers inside the HM \cite{Zhu2019} or by underlying 2D materials such as MoS$_2$ \cite{Xie2019}. In similar asymmetric systems, a new purely interfacial magnetoresistance effect, named spin-orbit magnetoresistance ascribed to Rashba spin-orbit interaction, has been predicted \cite{Grigoryan2014} and observed in ultrathin Cu-capped Pt/YIG systems producing an increase of the magnetoresistance \cite{zhou2018observation}. Moreover, ab-initio studies predicted that in-plane charge currents in Co/Pt and Pt/Cu interfaces can produce spin currents as large as those generated by spin Hall in bulk Pt \cite{Amin2016,Amin2016a, Amin2018}. Therefore, a more thorough experimental analysis of the interfacial effects is needed to understand the underlying phenomena and potentially obtain more efficient SOT devices. To this aim, Cu is a particularly promising material for applications given its low spin-orbit interaction and its bulk long spin diffusion ($\lambda_{sd}$).

In this work, we demonstrate experimentally a substantial increase of the SOT efficiency when introducing a Cu interlayer in a Pt/Cu($t_{Cu}$)/Co/Pt stack. The enhancement of SOT cannot be explained by bulk spin-Hall effect given the negligible spin Hall angle of Cu and considering the long spin diffusion in Cu, which can be as large as hundreds of nanometeres \cite{Sinova2015,Hoffmann2013}, making the Cu spacer between Pt and Co almost transparent to the spin currents generated by spin-Hall effect in Pt. The observed SOT can be thus solely ascribed to interfacial effects. The combined torques and spin magnetoresistance experiments as function of the Cu interlayer thickness demonstrate that spin currents generated at the interfaces are transduced directly to SOT. The magnitude of this effect must be proportional to the spin current density and independent of its polarization, so that our observation of a diminished spin magnetoresistance proves that spin memory loss drives the enhancement of the SOT. A simple drift diffusion model, based on Valet-Fert model \cite{Valet1993}, which accounts for spin currents and spin accumulation at the Cu/Co interface, reproduces satisfactorily the behaviour of the spin magnetoresistance for small and large Cu thickness when the spin memory loss is introduced. 

\section{methods}
The samples were prepared by DC sputtering onto thermally oxidized (300 nm) silicon substrates at Ar partial pressure of $8\cdot10^{-3}$ mbar. The base pressure of the chamber was $10^{-8}$ mbar. The film thicknesses were monitored in-situ using a quartz microbalance and confirmed by X-ray reflectivity experiments. The samples consisted in the following stack: Pt(5 nm)/Cu($t_{Cu}$)/Co(0.8 nm)/Pt(5 nm)/Ta(5 nm)//SiO$_2$(300 nm)/Si, with $t_{Cu}$ ranging from 0 to 3 nm. For $t_{Cu}$ = 0 nm (reference sample) we used a thicker Pt bottom layer to promote a net spin current that yields to a measurable SOT.

\begin{figure*}
    \includegraphics [trim=0 0 2cm 0,width=\textwidth]{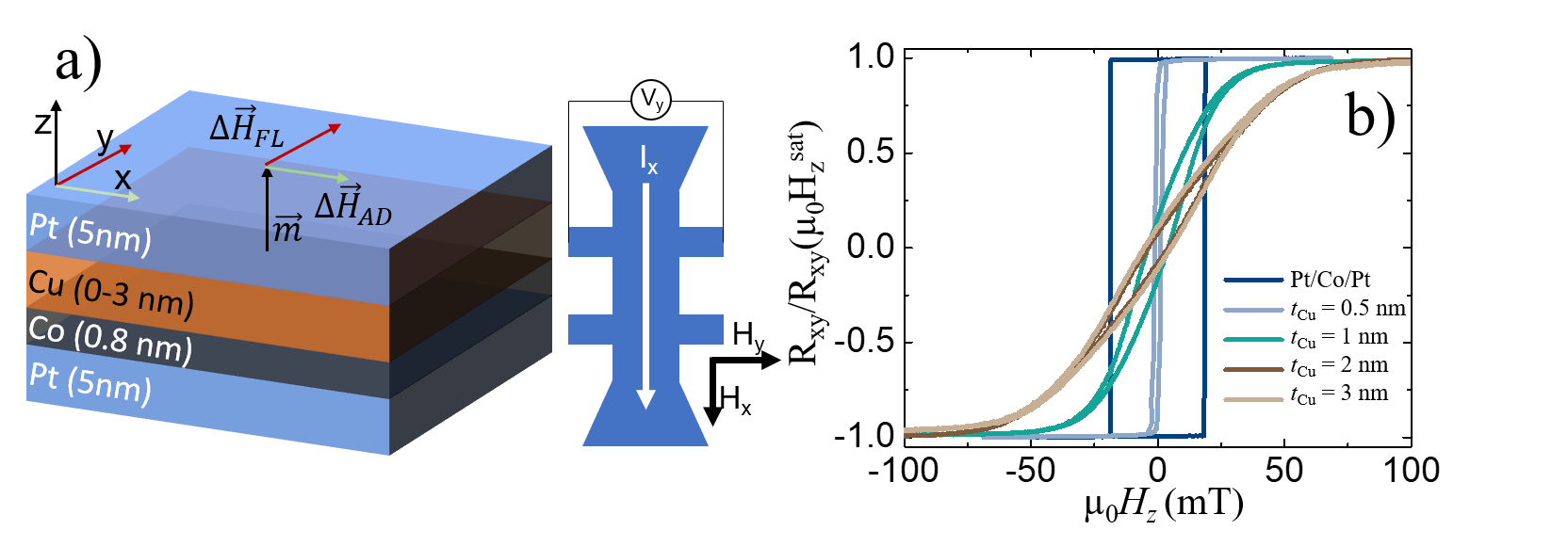}
    \caption{(Color online) {\bf Sample and Spin-orbit torque device schematics.} a) Schematic illustration of the structure and the Hall bar device for SOT measurements. b) Normalized room temperature transversal resistance as a function of the out of plane field for the structures with different Cu thickness interlayer. The increase of the thickness of the Cu interlayer produces a decrease in the effective perpendicular magnetic anisotropy.}
    \label{fig2}
\end{figure*}

The samples were processed using optical lithography to define Hall bar devices with channel width of 15 $\mu$m (photoresist AZ1512 and Ar milling etching were used). Ta(10 nm)/Cu(100 nm)/Pt(10 nm) electrical contacts were defined in a second optical lithography by DC sputtering.

Spin-orbit torque experiments were performed at room temperature as a function of a constant in-plane magnetic field, applied either along {\bf$\uvec{x}$} or {\bf$\uvec{y}$} directions, as shown in figure \ref{fig2}a. This technique, detailed in \cite{kim2013layer,PhysRevB.89.144425,Ruben2020}, allows for the estimation of antidamping-like (AD) and field-like (FL) torques induced to the FM. The torques are measured by analyzing the voltage response to a periodic electric current signal applied in the {\bf$\uvec{x}$} direction. The first and second harmonic components of the signal are obtained using Fourier transform of the measured voltages. The effective AD and FL torques ($\chi_{AD,FL}$) are obtained using the following equation:
\begin{equation}
\label{eq1}
\chi_{AD,FL}=-2\frac{\partial V^{2\omega}}{\partial H_{X,Y}}/\frac{\partial^2 V^{\omega}}{\partial H_{X,Y}^2}
\end{equation}


The contribution from the planar Hall effect is then considered and subtracted from the measured values of the effective SOT \cite{Woo2014}. 
The field dependent magnetization reversal pathways of the stacks were studied by means of Kerr magnetometry and anomalous Hall effect. The inclusion of the Cu interlayer gave rise to a reduction of the remanence magnetization, as shown in figure \ref{fig2}b, similarly to what was observed by Okabayashi et al.~\cite{Okabayashi2017}. Spin Hall Magnetoresistance (SMR) measurements were finally carried out at room temperature using a 1 T magnetic field and up to 5 mA in plane current. The sample was rotated in xy, xz and yz planes using a stepper motor.

Finally we have developed a drift-diffusion model based on Valet-Fert \cite{Valet1993} to account for the spin currents and spin accumulation in the system in order to further investigate the SMR dependence with $t_{Cu}$. Details of the model can be found in the supplementary information (section \ref{model}).

\section{Results and discussion}
\subsection{Spin-orbit torques in Pt/Co/Pt and Pt/Cu/Co/Pt}

\begin{figure*}
    \includegraphics [trim=0 0 0 0,width=\textwidth]{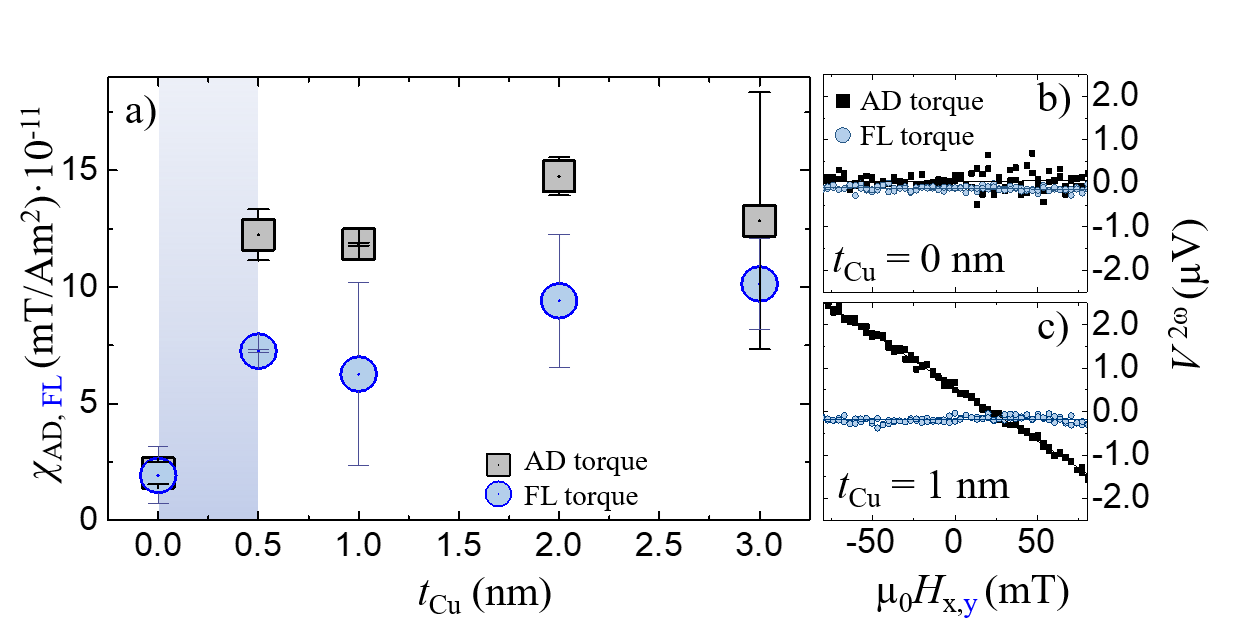}
    \caption{(Color online) {\bf Spin-orbit torque in Pt/Co/Cu($t_{Cu}$)/Pt stacks, as function of the Cu thickness $t_{Cu}$.} Antidamping-like torque increases six-fold when the Cu interlayer is introduced while field-like torque increases by a factor of 3. a) Evolution of the antidamping-like and field-like torques with $t_{Cu}$. b) Antidamping-like and field-like torques for the sample without a Cu interlayer and c) for $t_{Cu}=$1 nm. $\chi_{AD,FL}$ are calculated as shown in equation \ref{eq1} and then the contribution from planar Hall effect is substacted (see \cite{PhysRevB.89.144425,Woo2014}).}
    \label{fig3}
\end{figure*}

We measured the antidamping-like ($\chi_{AD}$) and field-like ($\chi_{FL}$) torque efficiencies as a function of the thickness of a Cu interlayer ($t_{Cu}$) ranging from 0 to 3 nm in the Pt(5)/Cu($t_{Cu}$)/Co(0.8)/Pt(5) stack (figure \ref{fig3}). In the case of absence of the Cu interlayer ($t_{Cu}$=0) we observe minimal values for both $\chi_{AD}$ and $\chi_{FL}$, as expected since the spin currents on top and bottom interfaces with Co are compensated (the direction of the spins point against each other due to the symmetry of the spin Hall effect), and therefore the observed SOT in both directions is small, i.e. $\chi_{AD,FL}<2\cdot10^{-11}\ {\rm mT/Am^2}$. Once a Cu interlayer is introduced ($t_{Cu}$=0.5 nm) we observe a six-fold increase of the antidamping-like torque up to $\chi_{AD}=12\pm 2\cdot10^{-11}\ {\rm mT/Am^2}$, whereas $\chi_{FL}$ increases by a factor of three. This value of $\chi_{AD}$ is comparable to the ones obtained in TaO$_x$/Co/Pt \cite{Woo2014,Ruben2020} or AlO$_x$/Co/Pt trilayers \cite{Garello2014,Lee2017} in which spin Hall effect in Pt is the main contribution. 

As the Cu interlayer thickness is further increased, $\chi_{AD}$ and $\chi_{FL}$ do not vary significantly, and the latter remains smaller than $\chi_{AD}$ in the whole thickness range. This behaviour suggests that the effect arises due to the inclusion of the Co/Cu interface and not due to spin accumulation changes nor to  spin-orbit interaction induced by proximity in the Cu layer. Moreover, since for large Cu thickness, both $\chi_{FL}$ and $\chi_{AD}$ are substantially unmodified, the observed enhancement of SOT at small thickness cannot be related with the Cu layer itself but it is rather due to a purely interfacial effect.
This may be caused either by the generation of spin currents at both Cu interfaces via spin filtering and spin precession mechanisms or due to a loss of spin current polarization by means of spin memory loss. Spin memory loss or interfacially generated spin currents cannot be distinguished by SOT measurements alone. Spin Hall magnetoresistance experiments provide the means to discern about the origin of this enhancement. In the case of spin memory loss, a decrease in the SMR is expected, while this is not the case for the interfacially generated spin currents. This will be discussed in the following.


\subsection{Spin-Hall magnetoresistance in Pt/Co/Pt and Pt/Cu/Co/Pt}
 SMR is the change of resistance in a FM/HM bilayer depending on the relative orientation of $\bf{m}$\ (reduced magnetization) and $\bf{j_c}$ (charge current density) vectors \cite{Chen2013,Chen2016,Kim2016}.
 This phenomenon is an effect of the spin current density generated in the HM impinging in the FM and being adsorbed by it, thus generating a second spin current in the HM that by inverse spin Hall effect is converted in an extra voltage \cite{Chen2016}. As the absorption depends on the FM magnetization direction and the spin current generated in this experiment is oriented along the {\bf$\uvec{y}$} direction, the SMR is expected to be maximum when the magnetization of FM is parallel to {\bf$\uvec{y}$}. The magnitude of the SMR is proportional to the squared spin Hall angle of the HM ($\theta_{SH}$, the efficiency of the spin current-voltage conversion) and to the spin mixing conductance.

\begin{figure*}
    \includegraphics[trim=0.5cm 0 0 1cm,width=\textwidth]{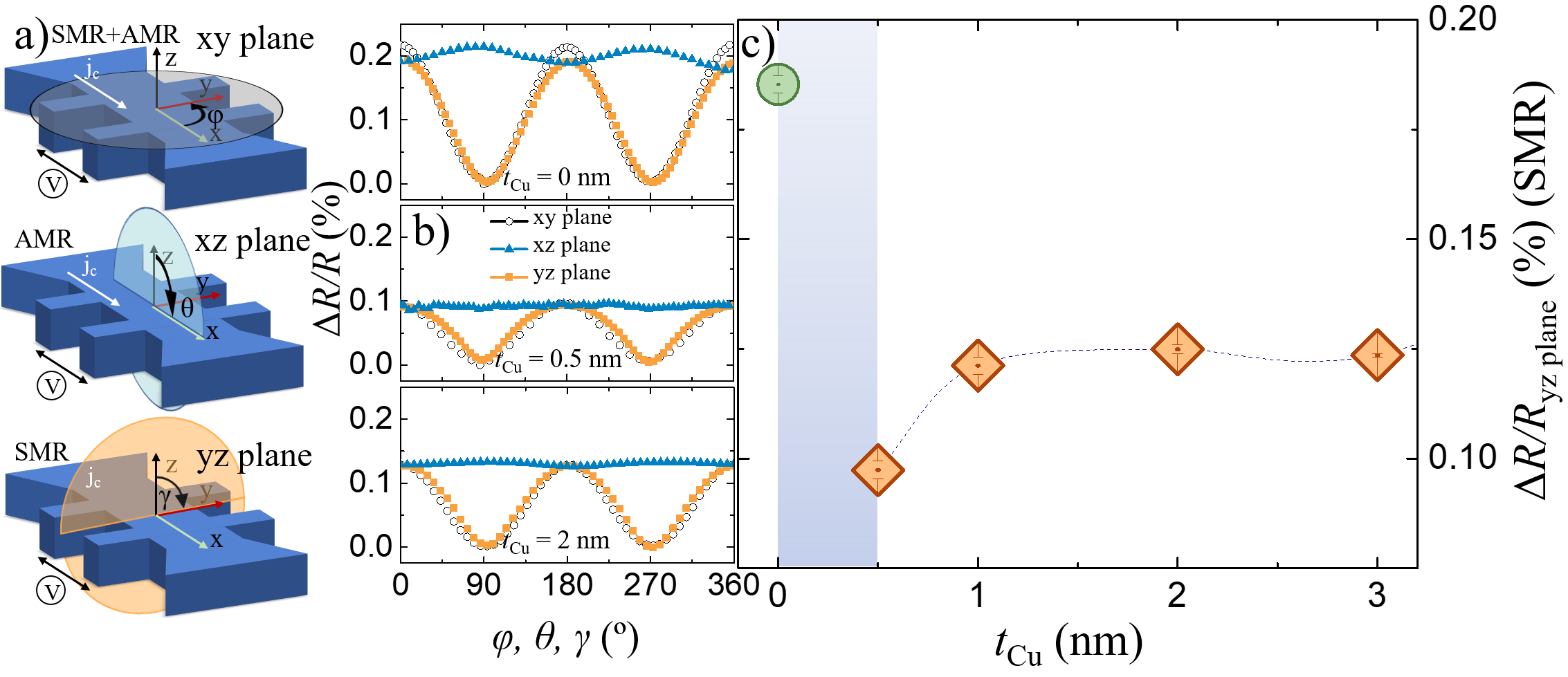}
    \caption{(Color online) {\bf Dependence of the Spin Hall magnetoresistance with Cu thickness.} a) Geometry of the measurement configurations. Spin-Hall magnetoresistance arises in the xy and yz plane configurations, while anisotropic magnetoresistance can be seen in the xz and xy plane configurations. The Spin-Hall magnetoresistance signal is observed in the yz plane. b) Resistance dependence on $\varphi$, $\theta$ and $\gamma$ angles at a magnetic field strength of $\mu_0$H = 1 T. c) Evolution of $\Delta R/R$ (Spin-Hall magnetoresistance) in the yz plane as a function of $t_{Cu}$. $\Delta$R represents the resistance difference between the maximum and minimum in the angular dependence and R is the minimum resistance. The line is a guide to the eye.}
    \label{fig4}
\end{figure*}

In the following, we analyse the Cu thickness behaviour of the SMR and anisotropic magnetoresistance. 
The geometry of the measurements is shown in figure \ref{fig4}a. The rotation in the xy plane combines anisotropic magnetoresistance and SMR, while the rotation in the zx and zy planes yield to the anisotropic magnetoresistance and SMR signals respectively. The dependence of the device resistance on the orientation of the magnetic field in the xy, xz and zy planes is shown in panel b of figure \ref{fig4}.

 A sizeable magnetoresistance appears in the three planes and the observed magnitude is in agreement with previous studies \cite{Avci2015}. The magnitude of the SMR is given by the ratio of the resistance difference between the maximum and minimum values and the minimum resistance ($SMR=(\Delta R/R)^{yz\ plane}=(R_{max}-R_{min})/R_{min}$) upon rotation of the magnetic field in the yz plane. The evolution of this magnitude with $t_{Cu}$ is shown in figure \ref{fig4}c. The plots in panels b and c show that the SMR magnitude decreases upon the introduction of the Cu layer and then slightly increases for larger thicknesses. Such a reduction of SMR is compatible with a reduction of the average spin current in the system, as a result of an enhancement of the spin memory loss.
The generation of spin currents would produce an enhancement of the spin accumulation at interfaces and therefore a small increment of the SMR. This has been observed experimentally when adding a thin W layer on top of a YIG/Pt system, even though W has opposite sign of the spin Hall angle when compared with Pt \cite{Luan2019}. We observe a decrease of the SMR of about 50\% when the Cu interlayer is placed. At larger $t_Cu$, SMR slightly increases and then remains approximately constant.

\begin{figure*}
    \includegraphics [trim={0 0 0 0},width=\textwidth]{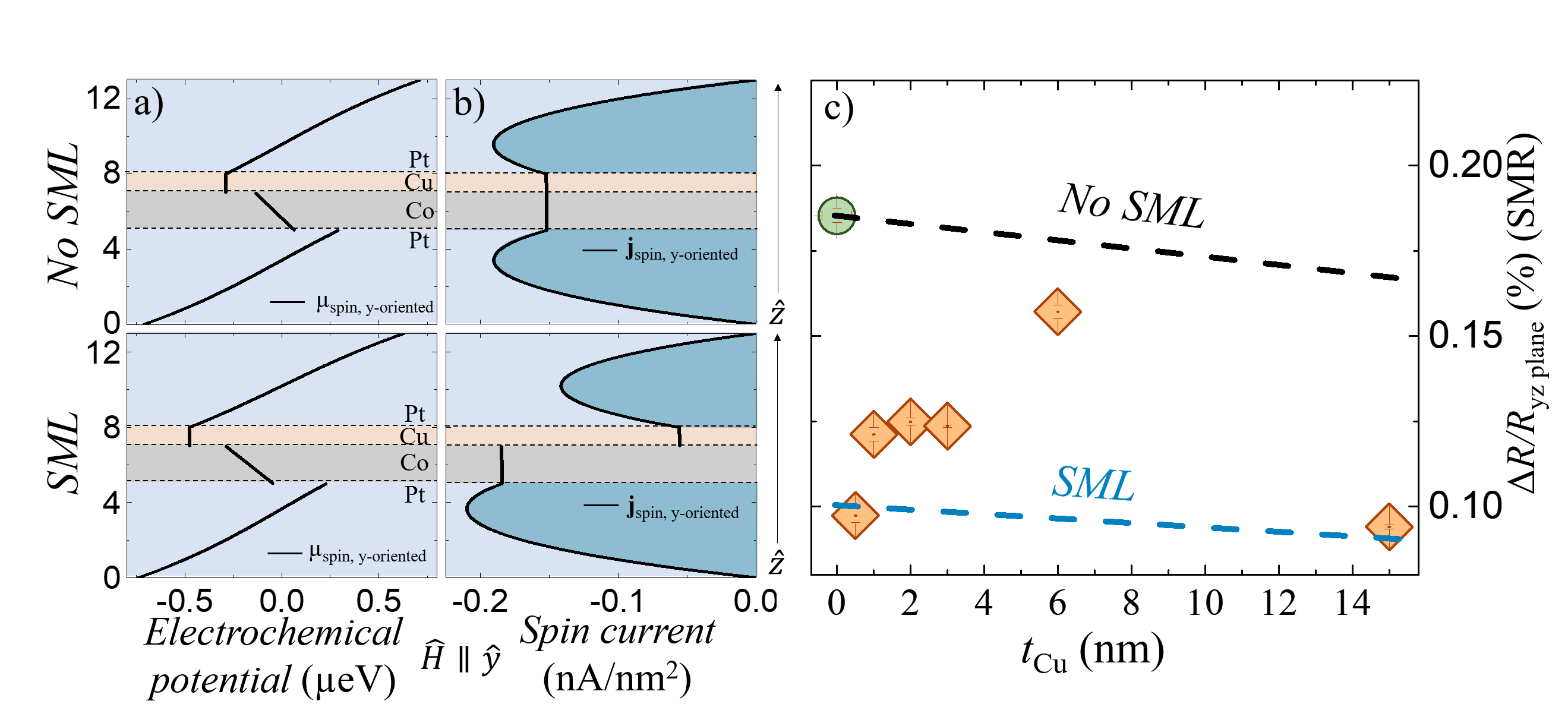}
    \caption{(Color online) {\bf Spin accumulation and spin Hall magnetoresistance.} a) Calculated spin electrochemical potentials in the system for $\mu_0H=1\ T$, $t_{Cu}=1$ nm, $t_{Co}=2$ nm and $t_{Pt}=5$ nm without considering spin memory loss and considering a spin memory loss of 0.7  \cite{Dolui2017} with $\bf{H}$ in the  $\bf{\widehat{y}}$ direction and b) their respective spin currents. Note the decrease of the spin current when spin memory loss is considered. c) Spin Hall magnetoresistance as a function of $t_{Cu}$ up to 15 nm. Dashed lines represent the modelled values of SMR as a function of $t_{Cu}$ normalizing the curve at the experimental value of $t_{Cu}=0\ nm$. The calculated SMR values are obtained by taking the difference of the electrochemical potential (spin accumulation) at the edges of the Pt layers (see supplementary information section \ref{model}). The modelled spin Hall magnetoresistance decreases linearly as $t_{Cu}$ increases due to the current shunting through Cu. The blue shadowed area in b) represents the spin current that contributes to the total spin Hall magnetoresistance, which is smaller in the system with spin memory loss.}
    \label{fig5}
\end{figure*}

This decrease in SMR when introducing the Cu interlayer suggests that the main mechanism involved in this behaviour is spin memory loss. To further understand the evolution of SMR and SOT with $t_{Cu}$, we have developed a drift-diffusion model to obtain the spin currents and accumulations in the system. This model is detailed in section \ref{model} of the supplementary information. Figures \ref{fig5}a and \ref{fig5}b show respectively the $\uvec{y}$-oriented spin accumulations and the out of plane $\uvec{y}$-oriented spin current densities for an external magnetic field along the $\uvec{y}$ direction. The SMR depends on the average spin current densities of both upper and bottom Pt layers (see supplementary information section \ref{SuppMetalLayers}), which gives rise to the spin Hall voltage via inverse Spin Hall effect. In order to further analyse the behaviour of SMR with $t_{Cu}$ we have extended its range up to 15 nm. These additional samples do not present perpendicular magnetic anisotropy and thus were not considered in SOT measurements.

The SMR extracted from the model (black dashed line in figure \ref{fig5}c) decreases linearly when $t_{Cu}$ increases. The spin memory loss reduces the spin current that flows through the Co/Cu interface (see figure \ref{fig5}b). For this calculation, we use the spin memory loss predicted by Dolui et al. \cite{Dolui2017}. Figure \ref{fig5}c shows that when the spin memory loss is accounted for, the total SMR decreases significantly (blue dashed line). This behaviour perfectly mimics our experimental observation when comparing both the Pt/Co/Pt and the Pt/Cu(0.5 nm)/Co/Pt sample, indicating that the SMR reduction is due to spin memory loss and is compatible with the both high and low $t_{Cu}$ regimes. At intermediate $t_{Cu}$ other additional effects must also be considered to describe completely the evolution of SMR with $t_{Cu}$. It is worth noting that although spin memory loss is typically considered to be a drawback in spintronic devices due to the detrimental overall magnetoresistance output, the accumulation of unpolarized spins at an interface may be used as a sink of spins to reinject in the device and augment the SOT efficiency.

\section{Conclusions}
 Spin-orbit interaction  can arise from bulk materials or at interfaces in which the low dimensionality breaks the structural inversion of symmetry. We have observed the emergence of an interfacially enabled increase of the spin-orbit torque when a Cu interlayer is inserted between Co and Pt in a symmetric Pt/Co/Pt trilayer, in which the effective spin-orbit torque is expected to vanish. The enhancement of SOT is accompanied by a reduction of the spin Hall magnetoresistance. We infer that spin memory loss in the Co/Cu and Cu/Pt interfaces is the main responsible for both enhanced SOT and reduction in the SMR at low and high Cu thickness regime. To understand the behaviour at intermediate Cu thickness we highlight the necessity to introduce other interfacial phenomena. The observed enhancement of the spin-orbit torque provides new insight on the interfacial nature of spin currents that can lead to develop more efficient spintronic devices using cheap, easy to fabricate and abundant materials like Cu.

\section{Acknowledgements}
We thank V.P. Amin for valuable discussions. We acknowledge D. Granados and A. Valera for their support in the optical lithography process to prepare the Hall bar devices. This research was supported by the Regional Government of Madrid through Project P2018/NMT-4321 (NANOMAGCOST-CM), by the Spanish Ministry of Economy and Competitiveness (MINECO) through Projects RTI2018-097895-B-C42 (FUN-SOC), FIS2016-78591-C3-1-R (SKYTRON), PGC2018-098613-B-C21 (SpOrQuMat), PGC2018-098265-B-C31 and  PCI2019-111867-2 (FLAG ERA 3 grant SOgraphMEM). JMD ackowledges support from MINECO through FPI program (BES-2017-080617). IMDEA Nanoscience is supported by the 'Severo Ochoa' Programme for Centres of Excellence in R\&D, MINECO  [grant number SEV-2016-0686].

\bibliography{references.bib}

\clearpage
\newpage
\section{Supplementary information}
\setcounter{figure}{0}
\renewcommand\thefigure{S\arabic{figure}}   

\subsection{Drift-diffusion model}
\label{model}
Drift-diffusion theory gives us an insight on the magnetic properties of metallic layers, such as spin Hall magnetoresistance and other phenomena. The study of these phenomena is based on the characterization of spin currents and spin electrochemical potentials in the different layers, via diffusion equations and interface phenomena \cite{Chen2013, Chen2016}. The description of spin currents and densities is different for ferromagnetic and normal metals

In ferromagnetic metals, electrons have different conductivity depending on their spin orientation relative to the direction $\uvec{m}$ of the magnetic field on the layer, either parallel $(\uparrow)$ or antiparallel $(\downarrow)$. Current densities associated to each orientation will be denoted respectively as ${\bf j}_{\uparrow F}$ and ${\bf j}_{\downarrow F}$. Assuming that all electrons are oriented along the direction of the magnetic field, the total charge current density ${\bf j}_{cF}$ and spin current density ${\bf j}_{sF}$ through the ferromagnetic layer are
\begin{equation}
{\bf j}_{cF} = {\bf j}_{\uparrow F} + {\bf j}_{\downarrow F}, \quad
{\bf j}_{sF} = {\bf j}_{\uparrow F} - {\bf j}_{\downarrow F}.
\end{equation}
Notice that spin currents are defined in units of Ampere; they can be converted to angular momentum currents multiplying by a factor $\hbar/2e$, with $e=-|e|$ the charge of an electron.

Current densities of parallel and antiparallel spins satisfy diffusion equations, with respect to their corresponding electrochemical potentials $\mu_{\uparrow F}$ and $\mu_{\downarrow F}$:
\begin{equation}
\label{SjF}
{\bf j}_{\varsigma F} = -\frac{\sigma_{\varsigma F}}{e} \nabla \mu_{\varsigma F}, \quad \varsigma = \uparrow, \downarrow,
\end{equation}
with $\sigma_{\uparrow F}$ and $\sigma_{\downarrow F}$ the conductivity of spins parallel and antiparallel to the magnetic field, respectively. The total conductivity of the ferromagnetic layer is therefore $\sigma_F = \sigma_{\uparrow F} + \sigma_{\downarrow F}$, and the conductance spin polarization is
\begin{equation}
P = \frac{\sigma_{\uparrow F} - \sigma_{\downarrow F}}{\sigma_{\uparrow F} + \sigma_{\downarrow F}}.
\end{equation}

Charge and spin accumulations in a ferromagnetic metal are represented by the following electroquemical potential:
\begin{equation}
\label{SmuF}
\mu_{cF} = \mu_{\uparrow F} + \mu_{\downarrow F}, \quad
\mu_{sF} = \mu_{\uparrow F} - \mu_{\downarrow F},
\end{equation}
which satisfy the following diffusion equations:
\begin{equation}
\label{SdiffF}
\nabla^2 \left( \mu_{cF} + \frac{1}{2} P \mu_{sF} \right) = 0, \quad
\nabla^2 \mu_{sF} = \frac{\mu_{sF}}{\lambda_F^2},
\end{equation}
with $\lambda_F$ the spin diffusion length in the FM.

\subsection{Metal layers}

In normal metals, due to arbitrary orientation of spins, the flow of electrons is described by a charge-current-density vector ${\bf j}_{cN}$ and a spin-current-density second-order tensor,
\begin{equation}
\label{SJtensor}
\mathfrak{J}_{sN} = ({\bf j}_{sNx}, {\bf j}_{sNy}, {\bf j}_{sNz})^T = ({\bf j}_{sN}^x, {\bf j}_{sN}^y, {\bf j}_{sN}^z),
\end{equation}
with row vectors ${\bf j}_{sN\iota}$ denoting $\iota$-oriented spin current densities, and column vectors ${\bf j}_{sN}^\iota$ denoting spin current densities flowing in the $\iota$ direction. In these layers, spin Hall effect (SHE) and inverse spin Hall effect (ISHE) may occur, as charge current generate orthogonal spin currents and viceversa. The SHE-generated $\iota$-oriented spin currents ${\bf j}_{sN\iota}^{SHE}$ and the ISHE-generated charge current ${\bf j}_{cN}^{ISHE}$ are described by the following relations:
\begin{equation}
\label{SHE}
\begin{split}
{\bf j}_{sN\iota}^{SHE} = \theta_{SH} \, \widehat{\bm{\iota}} \wedge {\bf j}_{cN}, \ \iota = x,y,z; \quad \\
{\bf j}_{cN}^{ISHE} = \theta_{SH} \sum_{\iota=x,y,z} \widehat{\bm{\iota}} \wedge {\bf j}_{sN\iota}, \quad
\end{split}
\end{equation}
where $\widehat{\bm{\iota}}$ denotes the unitary vector along the $\iota$ direction, and $\wedge$ denotes the vectorial product in $\mathbb{R}^3$. The spin Hall angle $\theta_{SH}$ is characteristic of each metal. In the following, it will be assumed to be small enough so that cubic and larger powers of $\theta_{SH}$ can be neglected.

Accumulation of electrons in a normal metal is described by a (scalar) charge electrochemical potential $\mu_{cN}$ and a (vector) spin electrochemical potential $\bm{\mu}_{sN} = (\mu_{sNx}, \mu_{sNy}, \mu_{sNz})$, with each component $\mu_{N\iota}$ representing the accumulation of spins oriented along the $\iota$ direction. The following equation is satisfied in the presence of SHE:
\begin{equation}
\begin{split}
\label{SjN}
{\bf j}_{cN} = -\frac{\sigma_N}{e} \nabla \mu_{cN}, \quad \\
{\bf j}_{sN\iota} = -\frac{\sigma_N}{2 e} \nabla \mu_{sN\iota} + {\bf j}_{sN\iota}^{SHE}.
\end{split}
\end{equation}
with the spin-Hall effect current density ${\bf j}_{sN\iota}^{SHE}$ induced by the charge current ${\bf j}_{cN}$ by \eqref{SHE} and where $\sigma_N$ is the conductivity of the metal. The electrochemical potentials $\bm{\mu}_N$ satisfy the following diffusion equations:
\begin{equation}
\label{SmuN}
\nabla^2 \mu_{cN} = 0, \quad
\nabla^2 \bm{\mu}_{sN} = \frac{\bm{\mu}_{sN}}{\lambda_N^2},
\end{equation}
The differences of the diffusion theory for normal metals with the one for ferromagnetic metals are the arbitrary orientation of spins, represented by the vectorial nature of $\bm{\mu}_{sN}$, and the decoupling of charge and spin accumulations when $P \to 0$.

The last ingredient in the description of the model is the analysis of the effect of the interface in spin current densities and accumulations. A simple description of the current along an interface between ferromagnetic and normal metals, as described in \cite{Chen2016}, is given by the following spin current density, flowing from the normal metal into the ferromagnetic metal:
\begin{equation}
\label{SjNF}
{\bf j}_s^{N|F} = (j_{\uparrow} - j_{\downarrow})\uvec{m} - \frac{G_r}{e} \uvec{m} \wedge (\uvec{m} \wedge \bm{\mu}_{sN}) - \frac{G_i}{e} \uvec{m} \wedge \bm{\mu}_{sN},
\end{equation}
with $G_{\uparrow \downarrow} = G_r + i G_i$ the spin-mixing conductance of the interface and

\begin{equation}
\begin{split}
\label{SjInt}
e r_\uparrow j_{\uparrow} = \left( \mu_{cN} - \mu_{cF} \right) + \frac{1}{2} \left( \uvec{m} \cdot \bm{\mu}_{sN} - \mu_{sF} \right), \quad \\
e r_\downarrow j_{\downarrow} = \left( \mu_{cN} - \mu_{cF} \right) - \frac{1}{2} \left( \uvec{m} \cdot \bm{\mu}_{sN} - \mu_{sF} \right),
\end{split}
\end{equation}

the flows of electrons parallel and antiparallel to the magnetic field along the interface, driven by the differences in charge and spin accumulations. Electrochemical potentials in \eqref{SjNF} and \eqref{SjInt} are evaluated at the corresponding sides of the interface.

\subsection{Multilayer system}
\label{SuppMetalLayers}

As indicated in the main text, the system of study is a multilayer system (see figure \ref{fig2}). It is thus possible to analyze spin accumulations in the layer by solving the drift-diffusion model for this system. The coordinate system chosen so that the z axis is orthogonal to the layer planes. An electric field is applied along the x axis, ${\bf E} = E\uvec{x}$, generating an in-plane charge current. Assuming xy-invariance of the system and that no charge current exists along the z axis, equations \eqref{SmuF} and \eqref{SdiffF} can be solved to determine the accumulations in ferromagnetic layers:
\begin{equation}
\begin{split}
\mu_{\uparrow F} = \mu_{cF} + \sigma_{\downarrow F} A e^{z/\lambda_F} + \sigma_{\downarrow F} B e^{-z/\lambda_F},\ \quad \\
\mu_{\downarrow F} = \mu_{cF} - \sigma_{\uparrow F} A e^{z/\lambda_F} - \sigma_{\uparrow F} B e^{-z/\lambda_F},
\end{split}
\end{equation}
with $A$ and $B$ integration constants determined by boundary condition, and $\mu_{cF}$ being constant along the z direction. Current densities are obtained by derivation, as in \eqref{SjF}:
\begin{equation}
j_{\uparrow F} = - j_{\downarrow F} = - \frac{\sigma_{\uparrow F} \sigma_{\downarrow F}}{e \lambda_F} \left( A e^{z/\lambda_F} - B e^{-z/\lambda_F} \right).
\end{equation}

In normal-metal layers, $\mu_{cN}$ is also constant along the z direction, while the spin accumulations are given by integration of \eqref{SmuN}:
\begin{equation}
\bm{\mu}_{sN} = {\bf C} e^{z/\lambda_N} + {\bf D} e^{-z/\lambda_N},
\end{equation}
with ${\bf C}$ and ${\bf D}$ constant vectors determined by boundary conditions. Spin current density only occurs along z direction; from \eqref{SJtensor}, \eqref{SHE}, and \eqref{SjN}, it is obtained that the spin current along z due to both diffusion and spin-Hall effect is
\begin{equation}
{\bf j}_{sN}^z = -\frac{\sigma_N}{2 e \lambda_N} \left({\bf C} e^{z/\lambda_N} - {\bf D} e^{-z/\lambda_N} \right) - \theta_{SH} \sigma_N E \uvec{y}.
\end{equation}

Thus, for a system of layers of known width, the description of spin currents and accumulations is completed by determining constants $\mu_{cF}$, $A$, and $B$ in ferromagnetic layers, and $\mu_{cN}$, ${\bf C}$, and ${\bf D}$ in normal-metal layers. These constants are determined by the following boundary conditions:
\begin{enumerate}
\item Spin currents are zero at the top and bottom boundaries of the multilayer system.
\item Spin density current through an interface between ferromagnetic and normal-metal layers is given by \eqref{SjNF}.
\item Spin density current through an interface between two normal-metal layers is conserved, as well as charge and spin electrochemical potentials.
\item Zero level for the charge electrochemical potential is arbitrarily set to zero at the bottom boundary of the multilayer system.
\end{enumerate}

The system of equations thus obtained for a multilayer system can be solved numerically. The results depend on the magnetization direction of the ferromagetic layers via the interface spin current densities. Figure XX presents solutions of the system for different orientations of the magnetization. The values for the physical constants have been taken as follows: for Co layers, $\sigma_F^{-1} = 46\, {\rm n\Omega \cdot m}$ \cite{Nguyen2014}, $\lambda_F = 30$ nm, $P=0,5$ \cite{Avci2015}; for Cu layers, $\sigma_N^{-1} = 6\, {\rm n\Omega \cdot m}$ \cite{Nguyen2014}, $\lambda_N = 300$ nm \cite{Villamor2013}, $\theta_{SH}=0$; for Pt layers, $\sigma_N^{-1} = 251\, {\rm n\Omega \cdot m}$, $\lambda_N = 3.43$ nm, $\theta_{SH}=5.3$ \%; for Co-Pt interfaces, $G_r = 5,5\cdot 10^{14}\,{\rm \Omega^{-1} m^{-2}}$ \cite{Tao2018}, $r_\uparrow = 0.70 \, {\rm f \Omega \cdot m^2}$, $r_\downarrow = 2.26 \, {\rm f \Omega \cdot m^2}$ \cite{Nguyen2014}; for Co-Cu interfaces, $G_r = 1.5\cdot 10^{13}\,{\rm \Omega^{-1} m^{-2}}$ \cite{Tokac2015}, $r_\uparrow = 0.24 \, {\rm f \Omega \cdot m^2}$, $r_\downarrow = 1.76 \, {\rm f \Omega \cdot m^2}$ \cite{Valet1993}. The imaginary part $G_i$ of the spin-mixing conductance of interfaces is negligible \cite{Chen2016, Jia2011}. The value of the resistivity of platinum that we obtain was measured using the Van der Pauw method on a 5 nm-thick Pt layer grown by DC sputtering. The resisitivity obtained is similar to the one reported by Sagasta et al. \cite{Sagasta2016} for the same range of thickness.

Boundary conditions 1 to 4 assume that spin current is completely transferred from one layer to another. If instead an spin memory loss (SML) effect is introduced, as described in the main text, then \eqref{SjNF} should be replaced by
\begin{equation}
{\bf j}_s^{N|F} = (1-\zeta_s)(j_{\uparrow} - j_{\downarrow})\uvec{m} - \frac{G_r}{e} \uvec{m} \wedge (\uvec{m} \wedge \bm{\mu}_{sN}) - \frac{G_i}{e} \uvec{m} \wedge \bm{\mu}_{sN},
\end{equation}
with $\zeta_s$ the absorption rate. Assuming an SML effect in the Co-Cu interface, with $\zeta_s = 70\%$, the resulting spin currents are represented in figure \ref{fig5} of the main text.

\subsection{Computation of SMR}

Spin current in Pt layers generates a charge current via ISHE, whose value can be computed from \eqref{SJtensor} and \eqref{SjN}. Hence, the total charge current through the layer induced by an applied electric field ${\bf E} = E \uvec{x}$ is
\begin{equation}
{\bf j}_{cN} = \sigma_N {\bf E} + \frac{\theta_{SH} \sigma_N}{2e} \left( \frac{\partial \mu_{sNy}}{\partial z} \, \uvec{x} - \frac{\partial \mu_{sNx}}{\partial z} \, \uvec{y} \right),
\end{equation}
and its average value for the thickness $t$ of the layer is
\begin{equation}
\label{SjCN-ISHE}
\overline{\bf j}_{cN} = \sigma_N {\bf E} - \frac{\theta_{SH} \sigma_N}{2e t} \left( \Delta \mu_{sNy} \, \uvec{x} -\Delta \mu_{sNx} \, \uvec{y} \right),
\end{equation}
with $\Delta \mu_{sN\iota}$ the difference of $\mu_{sN\iota}$ at the top of the layer minus its value at its bottom. Longitudinal and transverse charge currents appear due to SHE, effectively changing the resistivity of the multilayer system. In particular, if in the limit $\theta_{SH}\to 0$ the resistivity of the system is $\rho_0$ and the applied electric field generates a current density $j_0 \uvec{x}$, then the
spin-Hall effect causes a change $\Delta \rho_x$ in the longitudinal resistivity given by
\begin{equation}
\frac{\Delta \rho_x}{\rho_0} \simeq - \frac{\Delta j_x}{j_0},
\end{equation}
where $\Delta j_x$, the change in the longitudinal current density computed from the contribution of Pt layers through \eqref{SjCN-ISHE}, is assumed to be small.

\label{computationSMR}
\begin{figure*}[htbp]
    \includegraphics[width=15cm]{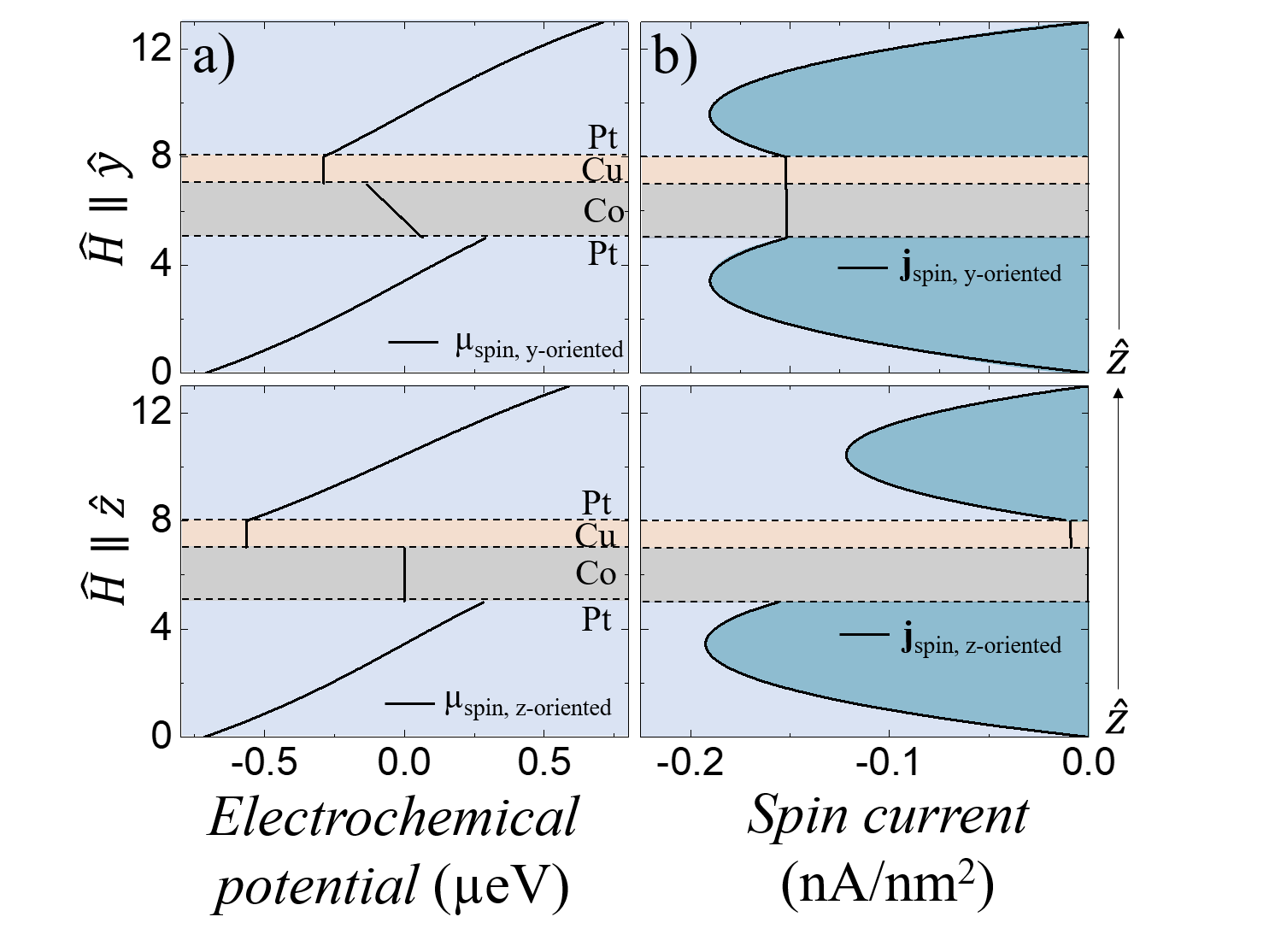}
    \caption{(Color online) {\bf Spin accumulations (a) and spin currents (b) for \textbf{H} in the \textbf{$\uvec{y}$} and \textbf{$\uvec{z}$} directions.} SMR can be computed directly considering the spin electrochemical potential difference in the spin-Hall active layers as shown in equation \ref{SjCN-ISHE}.}
    \label{figsupp1}
\end{figure*}

The computed value of $\Delta \rho_x$ depends on the direction $\uvec{m}$ of the magnetization of the ferromagnetic layers. Denoting by $\Delta \rho_x^{max}$ and $\Delta \rho_x^{min}$ its maximum and minimum values, it is immediate to compute the SMR, defined as the fraction of change in the resistance of the system, as
\begin{equation}
\frac{\Delta R}{R} = \frac{\Delta \rho_x^{max}-\Delta \rho_x^{min}}{\Delta \rho_x^{min}}.
\end{equation}

The minimum and maximum values correspond to \textbf{H} in the \textbf{$\uvec{z}$} and \textbf{$\uvec{y}$} directions respectively. Figure \ref{figsupp1} shows the values of $\mu_S$ and  $j_S$ through the {\bf$\uvec{z}$} direction in the system in study ($t_{Cu}=1\ nm$, $t_{Co}=2\ nm$ and $t_{Pt}= 5\ nm$).

\section{Evolution of SMR with Cu thickness}

\begin{figure*}
    \includegraphics [trim=0 1cm 0 0,width=\textwidth]{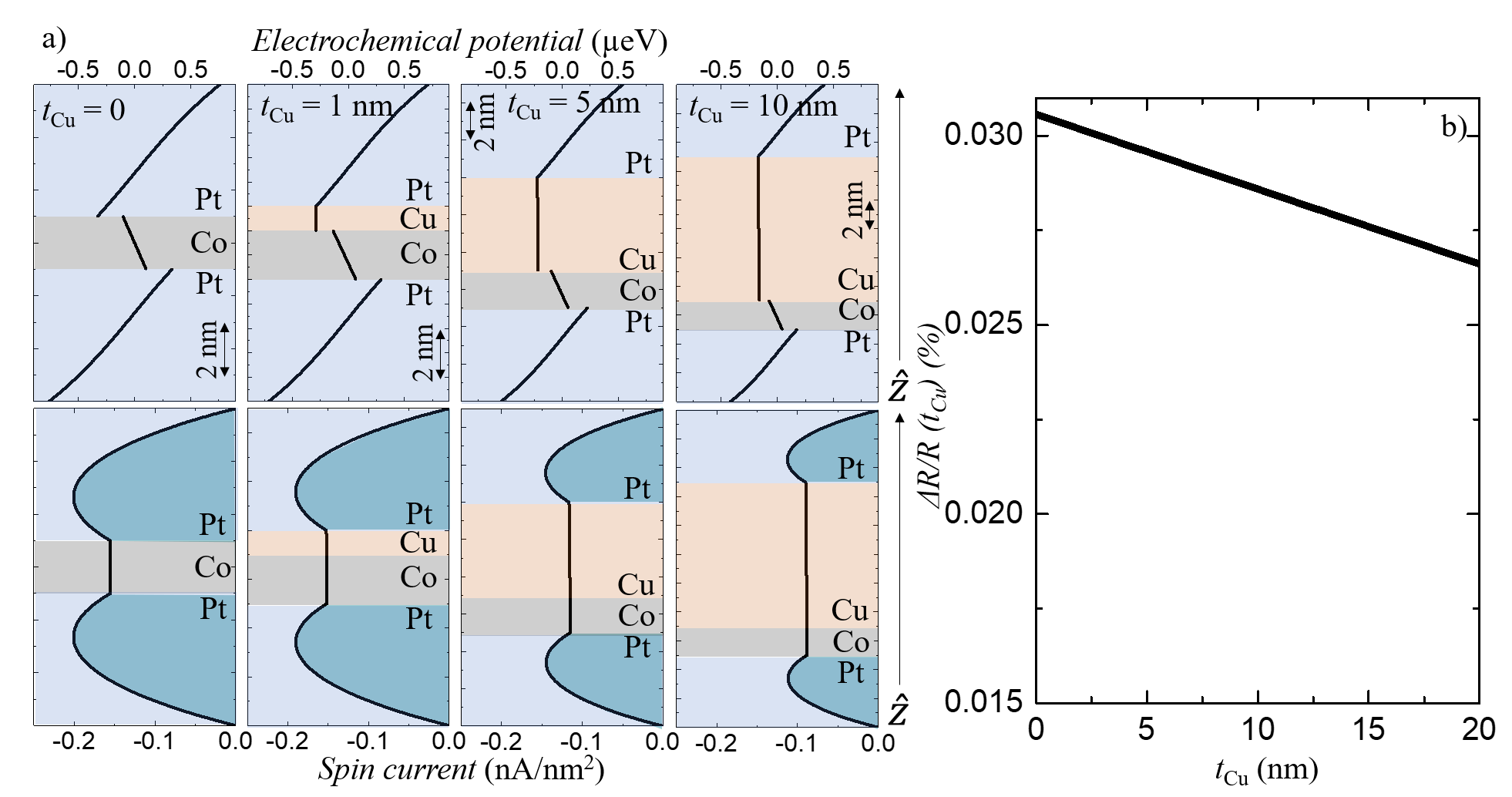}
    \caption{(Color online) {\bf Evolution of SMR with Cu thickness}. a) Electrochemical potentials for different values of $t_{Cu}$= 0, 1, 5 and 10 nm for \textbf{H} in the {\bf$\uvec{y}$} direction. The average spin current in the Pt layers decreases with increasing $t_{Cu}$, which results in a decrease of SMR. b) Dependence of $\Delta R/R$ as a function of $t_{Cu}$.}
    \label{figsupp2}
\end{figure*}

We can thus compute the values of $\frac{\Delta R}{R}$ as a function of $t_{Cu}$. We obtain a linear monotonic decrease when $t_{Cu}$ increases. The obtained values are shown in figure \ref{figsupp2} b).

\end{document}